# Strong Interaction of Cherenkov Radiation with Excitons in WSe$_2$ Crystals

Xuke Jiang[1], Masoud Taleb[1], Florian Diekmann[1], Kai Rossnagel[1,2], and Nahid Talebi[1]*

[1]Institute of Experimental and Applied Physics, Kiel University, 24098 Kiel, Germany

[2]Ruprecht Haensel Laboratory, Deutsches Elektronen-Synchrotron DESY, 22607 Hamburg, Germany

E-mail: talebi@physik.uni-kiel.de

The optical responses of semiconducting transition metal dichalcogenides are dominated by excitons. Being able to strongly interact with light and other materials excitations, excitons in semiconductors are prototypes for investigating many-particle and strong-field physics, including exciton-exciton, exciton-photon, and exciton-phonon interactions. Strong exciton-photon interactions, in particular, can lead to the emergence of exciton-polariton hybrid quasiparticles with peculiar characteristics, and a tendency toward macroscopic and spontaneous coherence. Normally, far-field and near-field optical spectroscopy techniques are used to investigate exciton-photon interactions. Here, we demonstrate that the radiation generated by moving electrons in transition metal dichalcogenides, namely Cherenkov radiation, can strongly interact with excitons. We investigate the coherence properties and spectral signatures of exciton-photon interactions in TMDC bulk crystals, using cathodoluminescence spectroscopy. Our findings lay the ground for cathodoluminescence spectroscopy and in particular electron-beam techniques as probes of exciton-polariton spontaneous coherence in semiconductors, beyond the well-known plasmonic investigations.

**Introduction**

Van der Waals materials have peculiar optical properties[1,2], ranging from topological surface excitations and Dirac plasmons in the THz range[3,4] and phonon polaritons in the IR range[5-7], to the excitation of exciton-polaritons in the visible range[8-10]. A group of van der Waals materials, namely, semiconducting transition metal dichalcogenides (TMDCs), can support a variety of exciton excitations, including bright excitons distinguished by spin-orbit interactions[11], dark excitons[12], bi-excitons[13], and charged excitons known as trions[14]. In particular, group VI TMDCs $MX_2$ with M = Mo, W and X = S, Se, and Te exhibit fascinating excitonic structures with unique opportunities to optically manipulate spin and valley states. In multilayers of these materials, the electronic structure demonstrates an indirect bandgap, whereas in monolayers, the bandgap becomes direct. Even in the bulk, however, excitons can strongly interact with photons, where for enhancing light-matter interactions, normally optical microcavities are used[15-17].

Here, we propose and experimentally investigate another scenario for strong exciton-photon interactions, mediated by the radiation from the electron beams propagating in bulk TMDCs, namely, Cherenkov radiation (CR)[18].

CR is coherent radiation emitted from electron beams in materials with refractive indices larger than a critical value. Similar to acoustic shock waves and sonic booms, the radiation caused by the disturbance (here the propagating particle) constructively build up in certain directions (see Fig. 1), with its emission angle depending on the speed of the particle propagating in the medium. Therefore, analyzing CR, i.e., its time of flight and direction, provides accurate means for particle identification[19,20]. Here, we argue that CR in exciton-active TMDCs particularly leads to another degree of freedom for an accurate analysis of particle velocities, thanks to the strong interaction of the CR with excitons in these materials. We first provide the theory of CR in bulk $WSe_2$ crystals and analyze the anticrossing effect caused by strong exciton-photon interactions versus the electron velocity. It will be shown that the CR angle is larger than the critical angle for the total internal reflection in $WSe_2$ planar crystals; thus CR can be captured inside thick crystalline films and cause Fabry-Perot resonances, which further strengthen the CR-exciton interactions. Our analysis could pave the way toward realizing novel kinds of scintillators based on strong-coupling effects.

## Results

**Cherenkov radiation in bulk WSe$_2$ and boundary effects.**

We use a vector-potential approach to calculate the radiation from electron beams in bulk materials and thin films, as well as the intensities of momentum-resolved electron energy-loss spectroscopy (MREELS)[21,22]. The current density of a moving electron with speed $v_{el}$ along the z-axis is given by $J_z(\vec{r},t) = -ev_{el}\delta(x)\delta(y)\delta(z-v_{el}t)$. The vector potential component $A_z$ is derived by solving the inhomogeneous Helmholtz equation $\vec{\nabla}^2 A_z(\vec{r},\omega) + \varepsilon_r(\omega)k_0^2 A_z(\vec{r},\omega) = -\mu_0 \tilde{J}_z(\vec{r},\omega)$ in a bulk medium with a permittivity of $\varepsilon_r$ where $\tilde{J}_z$ is the current density in the frequency domain, as

$$A_z(\vec{r},\omega) = \frac{1}{4\pi^2}\int_{-\infty}^{+\infty}\int_{-\infty}^{+\infty}\tilde{A}_z(k_x,k_y,z;\omega)e^{-ik_y y}e^{-ik_x x}dk_x dk_y, \quad (1)$$

with $\tilde{A}_z$ given by

$$\tilde{A}_z = e\mu_0\mu_r \frac{1}{\varepsilon_r(\omega)k_0^2 - k_x^2 - k_y^2 - (\omega/v_{el})^2}\exp\left(i\frac{\omega}{v_{el}}z\right). \quad (2)$$

From the solutions to the vector potential, the field components are calculated as [23]

$$\vec{E}(\vec{r},\omega) = -i\omega\vec{A}(\vec{r},\omega) + \frac{[\hat{\varepsilon}_r(\omega)]^{-1}}{i\omega\varepsilon_0\mu_0}\vec{\nabla}\left(\vec{\nabla}\cdot\vec{A}(\vec{r},\omega)\right) \quad (3)$$

for the electric field and

$$\vec{H}(\vec{r},\omega) = \frac{1}{\mu_0}\vec{\nabla}\times\vec{A}(\vec{r},\omega) \quad (4)$$

for the magnetic field.

Using the Poynting theorem, the power radiated to the far-field zone along the z-axis is calculated as $P_z = \frac{1}{2}\text{Re}\int dx \int dy \left(E_x H_y^* - E_y H_x^*\right) = \int dk_x \int dk_y \tilde{p}_z\left(\omega;k_x,k_y\right)$. Here, $\tilde{p}_z\left(\omega;k_x,k_y\right) = \frac{1}{2}\text{Re}\left(\tilde{E}_x \tilde{H}_y^* - \tilde{E}_y \tilde{H}_x^*\right)$ is the radiated power calculated in reciprocal space. As a result of the interaction of the moving electron within a bulk material, $\tilde{p}_z$ is given by

$$\tilde{p}_z\left(\omega;k_x,k_y\right) = \frac{e^2}{8\pi^2 \varepsilon_0 v_{el}} \text{Re} \frac{k_x^2 + k_y^2}{\varepsilon_r} \frac{1}{\left|\varepsilon_r k_0^2 - k_x^2 - k_y^2 - \left(\omega/v_{el}\right)^2\right|^2} . \tag{5}$$

The EELS integral will be given by the action integral[24]:

$$\Gamma^{EELS}(\omega) = \frac{-1}{\hbar\omega}\text{Re}\iiint_v \vec{E}(\vec{r},\omega)\cdot\vec{J}^*(\vec{r},\omega)dv$$

$$= \frac{e}{2\pi\hbar\omega}\text{Re}\iint_{k_x,k_y} dk_x dk_y \int_{z=-\infty}^{+\infty} dz\, \tilde{E}_z\left(k_x,k_y;z;\omega\right)e^{-i\frac{\omega}{v_{el}}z}, \tag{6}$$

and the MREEL spectrum in diffraction is defined as

$$\Gamma_k^{EELS}\left(k_x,k_y;\omega\right) = \frac{e}{2\pi\hbar\omega}\int_{z=-\infty}^{+\infty} dz\, \tilde{E}_z\left(k_x,k_y;z;\omega\right)e^{-i\frac{\omega}{v_{el}}z}. \tag{7}$$

By combining this with the electric field calculated using eq. (3), the MREEL spectrum is obtained as

$$\frac{d\Gamma^{EELS}\left(\omega;k_x,k_y\right)}{dz} = \frac{e^2}{4\pi^2\hbar\omega^2\varepsilon_0}\text{Im}\left\{\frac{1}{\varepsilon_r}\frac{\left(\varepsilon_r k_0^2 - \left(\omega/v_{el}\right)^2\right)}{\varepsilon_r k_0^2 - k_x^2 - k_y^2 - \left(\omega/v_{el}\right)^2}\right\}. \tag{8}$$

Here, $\vec{k} = \left(k_x,k_y,k_z=\omega/c\right)$ is the wave vector of the emitted light in reciprocal space, $e$ is the electron charge, and $k_0 = \omega/c$ and $\varepsilon_0$ are the free-space wavenumber and permittivity, respectively.

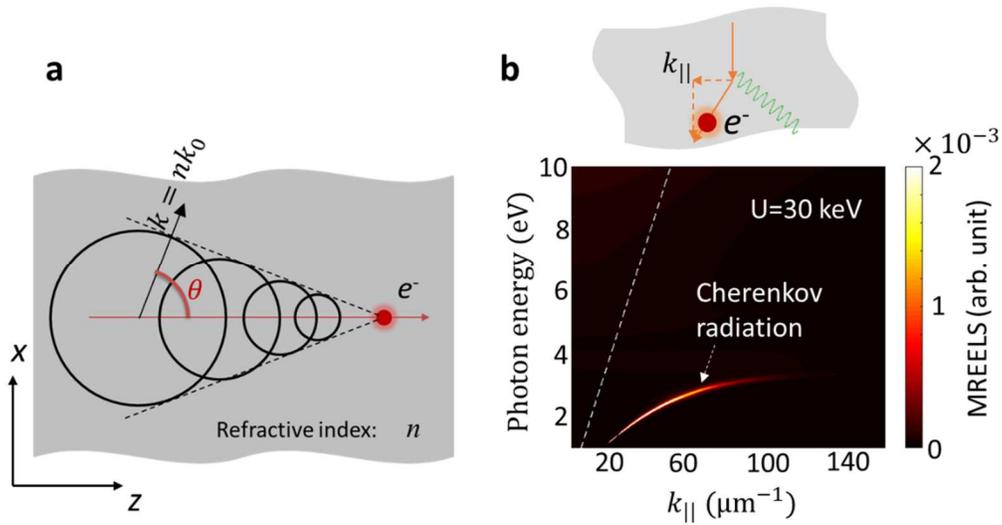

**Fig. 1** (a) Schemtaic illustration of Cherenkov radiation in a bulk material. (b) Dispersion of Cherenkov radiation as obtained by MREELS theory.

The results of MREELS calculations, performed for electrons moving inside silicon, demonstrate a sharp energy-dependent peak, that is associated with the CR emitted from the electrons. Evidently, CR dispersion is positioned outside the light cone (Fig. 1b). CR in semiconductors and dielectrics can contribute to the overall electron energy-loss spectra and therefore poses a limit on the accurateness of the models used for Kramers-Krönig analysis to retrieve the materials dielectric function using EELS[25].

Normally, a simple geometrical optics argument is exploited to model the Cherenkov radiation and its propagation direction with respect to the electron trajectory inside the material. This argument, which is based on the phase-matching condition between the electron current density and the CR, thus implies $\omega v_{el}^{-1} = nk_0 \cos\theta$ (see fig. 1a), demonstrating the strong dependence of the CR dispersion on the velocity of the electron.

CR generated by an electron moving inside a WSe$_2$ crystal can strongly interact with excitons in this material. Due to strong spin-orbit interactions, the degeneracy of the $K$ and $-K$ points in the Brillouin zone is split by a large energy of about 400 meV, that results also in fine structures for bandgap excitons, referred to as A excitons (excited at an energy of E = 1.68 eV) and B excitons (E = 2.1 eV).

CR is generated only above a critical electron kinetic energy of $U = 12$ keV ($v_{el} = 0.213\,c$) (see Fig. 2a). Therefore, at electron energies below this critical value, only excitonic absorption peaks are apparent in the MREELS maps. However, by increasing the kinetic energy of the electron to above $U = 12$ keV, CR emission is excited at energies lower than the A exciton energy. The excited CR strongly interacts with the A excitons, and results in exciton-polaritons propagating in the bulk of the material. The electron kinetic energy further directly affects the coupling strength and the related energy splitting, as well as the excitation efficiency of lower polariton (LP) and upper polariton (UP) branches, since it alters the CR dispersion and its spectral weight.

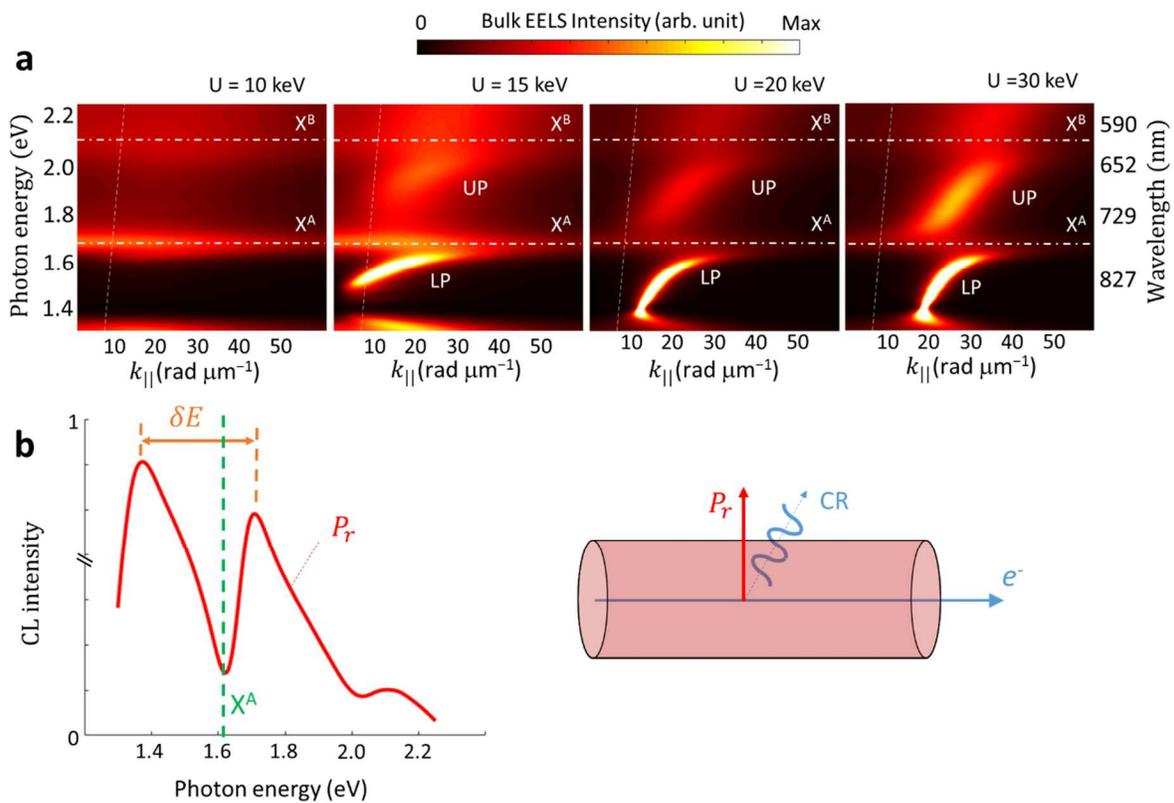

**Fig. 2** (a) Calculated MREELS maps revealing the interaction of a swift electron in bulk WSe$_2$ at different electron energies. (b) Calculated CR power spectra for the power flowing normal to the electron trajectory.

The radiated electromagnetic power flow, modelled by the Poynting vector, can be decomposed into components vertical ($P_r$) and parallel ($P_z$) to the propagation direction of the electron. We notice that the strong coupling effect and energy splitting are well represented in the $P_r$ component, where a dip at the A exciton energy is observed – instead of a peak, as apparent for example in photoluminescence spectra of WSe$_2$ crystals[26], where no CR is excited (see Fig. 2b). Indeed, the exciton absorption peak is split into two broader peaks with an energy splitting on the order of $\delta E = 0.335$ eV. A weaker interaction with the B exciton is observed as well, indicated again by a dip at an energy of $E = 2.1$ eV.

Although the strong CR-exciton interactions, and the CR emission in general, ideally occur in the bulk of the crystals, surfaces and interfaces provide real experimental environments. The emitted CR and the excited exciton-polaritons can thus interact with the boundaries. In particular, for electrons at a kinetic energy of 30 keV, the CR emission angle ($\theta_{CR}$) is considerably larger than

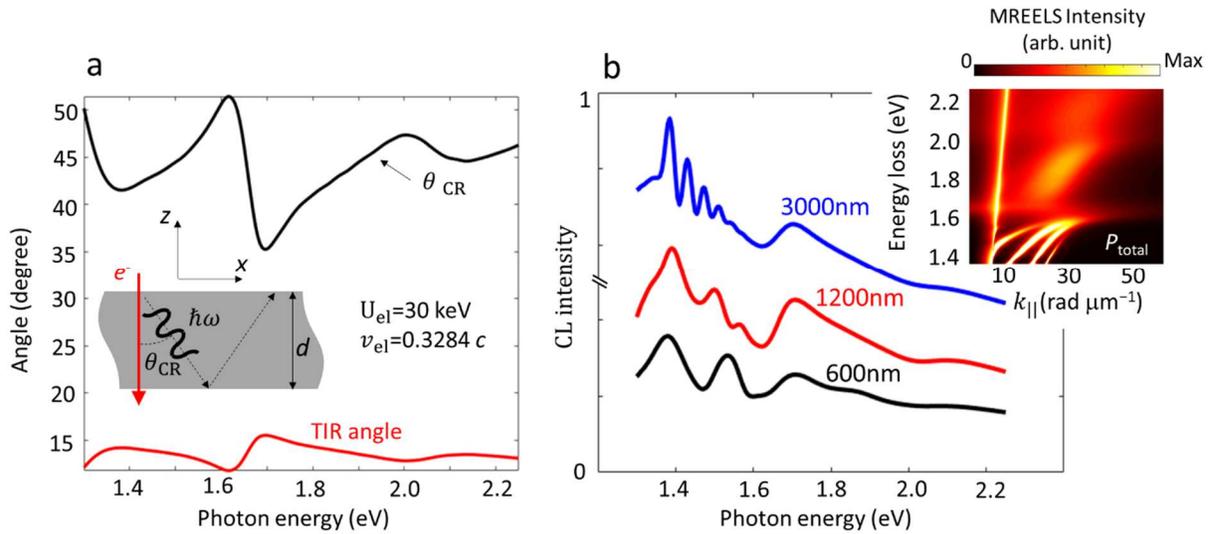

**Fig. 3** (a) Angle of the emitted Cherenkov radiation with respect to the electron trajectory, compared to the critical angle for total internal reflection (TIR), explaining why the Cherenkov radiation is captured within the thin film. (b) CL emission spectra $P = P_{+z} + P_r$ for a film with the depicted thicknesses, where the spectral fringes caused by the internal Fabry-Perot interferences are obvious. Inset: MREELS intensity for a film with a thickness of $d = 600$nm.

the critical angle for the total internal reflection (TIR) of the beams at the planar boundaries. Therefore, the CR emission, particularly the LP branch, is confined to the inside of the films allowing for Fabry-Perot resonances due to multiple reflections at the boundaries. This effect is readily captured in the calculated MREELS map (Fig. 3b, inset), where the dispersion of each individual mode is obviously traceable. The spectral fringes caused by the Fabry-Perot resonance effect are observed in the power flow of the emitted light along the axis normal to the interface ($P_z$), where the strong coupling effect is better visible in the $P_r$ component, and the associated dips at the A and B exciton energies are still present.

**Experimental Results.**

High-quality 2H-WSe$_2$ single crystals were grown using a standard chemical vapor transport technique, they were directly used throughout this study without further treatments.

A field emission scanning electron microscope (FE-SEM) (Zeiss SIGMA) operating in 5 - 30 kV was used for electron-probe measurements. Cathodoluminescence investigations were carried out utilizing detectors (Delmic B.V) attached to the FE-SEM setup. WSe$_2$ specimen surfaces were scanned and excited by a beam current of 14 nA causing CL radiation. The emitted radiation was collected by an off-axis silver parabolic mirror (acceptance angle: 1.46π sr, dwell time: 400 μs) which was installed above the specimen in a focal point distance of 0.5 mm. The collected radiation was directed to a dispersive grating and a CCD camera.

Fig. 4a shows CL spectra acquired from a thin multilayered WSe$_2$ structure. As theoretically predicted, the CL spectrum shows a dip – instead of a peak – at the A exciton energy of E=1.61 eV. In addition, we observe an energy splitting on the order of $\delta E = 0.35$ eV, in good agreement with the theoretically predicted value of $\delta E = 0.335$ eV. We observe an additional peak at $E = 1.36$ eV, which we attribute to the rather complicated boundary effects in this structure.

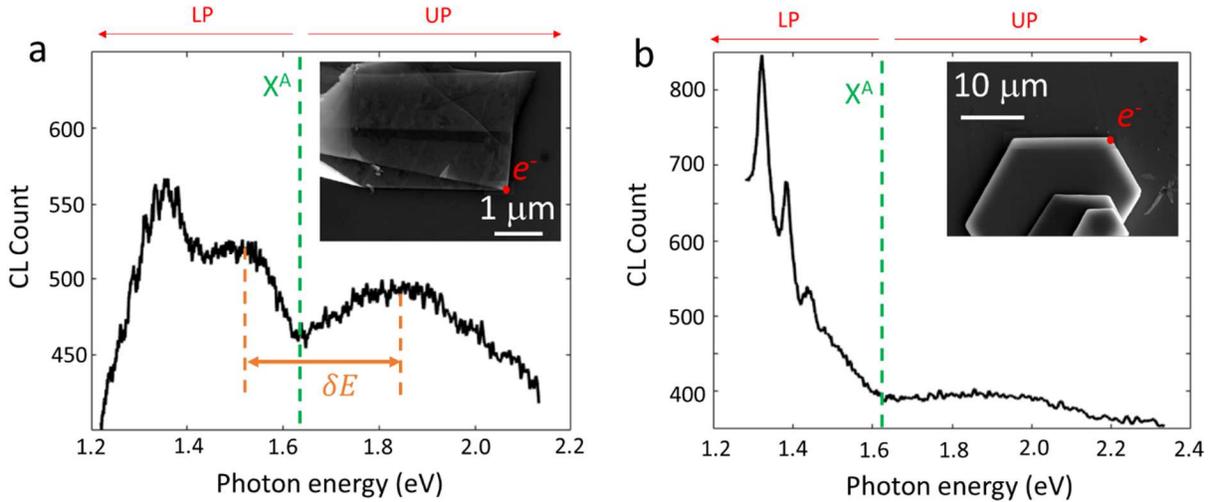

**Fig. 4** Cathodoluminescence spectra for an electron at an energy of 30 keV interacting with a (a) multi-layered WSe₂ structure and (b) thick WSe₂ crystal. In both cases, the strong interaction of Cherenkov radiation with excitons is indicated by a dip in the spectra at the A exciton energy.

For crystalline WSe₂ flakes with thicknesses larger than the wavelength of the CR emission, Fabry-Perot resonances inside the films are perfectly captured in the measured CL spectrum (see Fig. 4b). From comparison with theory, we anticipate a thickness of $d = 2$ µm for the crystal, which agrees well with the thickness measurement based on SEM images recorded at tilted views. Again, a dip at the exciton energy signifies a strong CR-exciton coupling effect. Both in the theoretical and experimental results, we do not observe any spectral fine structure in the UP branch, in contrast to the LP branch. We attribute this behavior to the larger attenuation constant of the upper polaritons, as well as the lower oscillator strength attributed to the B excitons, compared to the A excitons. Indeed, A excitons couple more efficiently to the CR emission and in general to the light, compared to B excitons.

By precisely tuning the kinetic energy of the electron in our SEM setup, we indeed observe that the CR emission is detetcted only at acceleration voltages higher than $U = 10$ kV (see Fig. 5). Moreover, by increasing the acceleration voltage, the CR emission and thus the overall intensity of the CL spectrum are strongly enhanced. The dip at the A exciton energy is an omnipresent effect in all the acquired CL spectra above 10kV acceleration voltage and signifies that indeed the strong-coupling effect is induced by CR emission.

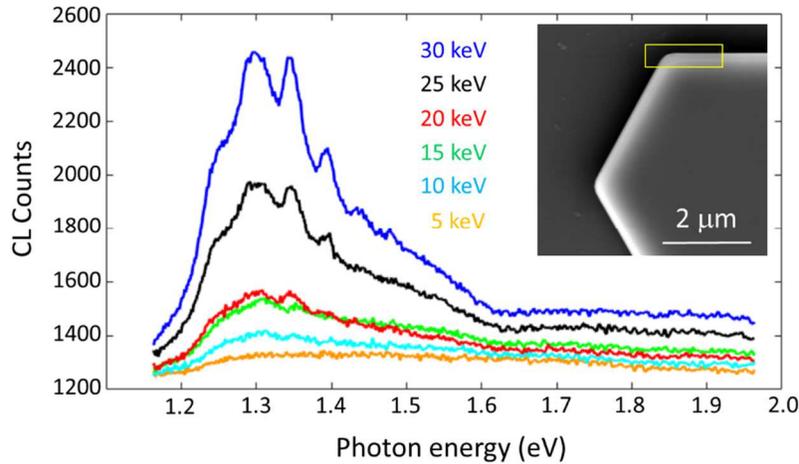

**Fig. 5** Cathodoluminescence spectra at different electron energies showing a strong spectral contribution of the captured CR at higher electron energies.

Here, we have only discussed the strong coupling of the CR emission with excitons in thick $WSe_2$ structures. However, in passing we emphasize that, indeed, the photonic modes of thinner TMDC films can also strongly interact with excitons, as shown by scanning near-field optical microscopy[27]. Moreover, the associated spatial interference fringes and spectral features can be fully retrieved by cathodoluminescence spectroscopy, as will be reported in another contribution.

In summary, we hypothesized and theoretically and experimentally investigated the strong interaction of CR with excitons in exciton-active, semiconducting TMDCs. Although our studies considered only single-crystalline $WSe_2$ structures, we expect the same phenomena to occur in other semiconductors with room-temperature excitons. We further demonstrated that CR can be captured in slabs thicker than the CR wavelength and can cause Fabry-Perot resonances. These Fabry-Perot resonances emerge as fine structure in the acquired CL spectra and demonstrate the fully coherent process of CR emission, happening due to its phase-matched excitation nature, with respect to the evanescent field accompanying the moving electron beam inside the material.

**Acknowledgement**

This project has received funding from the European Research Council (ERC) under the European Union's Horizon 2020 research and innovation programme, Grant Agreements No. 802130 (Kiel, NanoBeam) and Grant Agreements No. 101017720 (EBEAM). Financial supports from Deutsche Forschungsgemeinschaft under the Art. 91 b GG Grant Agreement No. 447330010 and Grant Agreement No. 440395346 is acknowledged.